\documentstyle[prb,aps]{revtex}

\begin{document}
\wideabs { \draft
\title{Detailed analysis of the mean diameter and diameter
distribution of single wall carbon nanotubes from their optical
response}

\author{X. Liu,$^1$ T. Pichler,$^{1,2}$ M. Knupfer,$^1$ M.S.
Golden,$^3$ J. Fink$^1$,  H. Kataura,$^4$ and Y. Achiba$^4$}

\address{$^1$ Institute for Solid State and Materials
Research Dresden, P.O. Box 270016, D-01171 Dresden, Germany\\
$^2$ Institut f\"ur Materialphysik, Universit\"at Wien,
Strudlhofgasse 4, A-1090 Wien, Austria. \\
$^3$ Van der Waals-Zeeman Institute, University of Amsterdam, Valckenierstraat
65, NL - 1018 XE Amsterdam, The Netherlands. \\
 $^4$ Faculty of Science, Tokyo
Metropolitan University, 1-1 Minami-Ohsawa, Hachioji, Tokyo 192-0397, Japan  }

\date{\today}

\maketitle
\begin{abstract}

We report a detailed analysis of the optical properties of single wall carbon
nanotubes with different mean diameters as produced by laser ablation. From a
combined study of optical absorption, high resolution electron energy-loss
spectroscopy in transmission and tight binding calculations we were able to
accurately determine the mean diameter and diameter distribution in bulk SWCNT
samples. In general, the absorption response can be well described assuming a
Gaussian distribution of nanotube diameters and the predicted inverse
proportionality between the nanotube diameter and the energy of the absorption
features. A detailed simulation enabled not only a determination of the mean
diameter of the nanotubes, but also gives insight into the chirality
distribution of the nanotubes. The best agreement between the simulation and
experiment is observed when only nanotubes within 15$^\circ$ of the armchair
axis are considered. The mean diameters and diameter distributions from the
optical simulations are in very good agreement with the values derived from
other bulk sensitive methods such as electron diffraction, x-ray diffraction
and Raman scattering.
\end{abstract}
 }
\newpage

{\section {Introduction}}

\par
Since the discovery of carbon nanotubes \cite{ijma}, a great deal of attention
has been focused on this entirely new class of nanoscale materials. Due to
their unusual geometry, their structural and electronic properties, these
carbon nanostructures are viewed as promising building blocks for molecular
electronics \cite{dekker}. In particular, single-wall carbon nanotubes (SWCNT)
are currently being intensively investigated worldwide since they possess
unique and intriguing electronic properties, being either semiconducting or
metallic depending on their geometrical structure defined by their chirality
\cite{hamada,saito}. After an effective production method of SWCNT was
discovered \cite{guo}, a huge number of investigations were initiated. However,
in all production methods available today (laser desorption \cite{guo}, carbon
arc method \cite{arc}, HPCO (high pressure CO decomposition)\cite{co}, and
chemical vapor deposition (CVD) \cite{cvd_dai}) the produced SWCNT are formed
as a polydisperse mixture with various diameters and chiralities. Although
claims have been made for the formation of a SWCNT lattice with only one type
of (10,10) nanotubes\cite{gimcevsky}, generally speaking the process of
synthesizing nanotubes of only a single diameter and chirality is still beyond
our reach.
\par
A first step in a systematic approach towards improved selectivity during the SWCNT production process is a
feedback of information coming from a reliable characterization of the mean diameter and diameter distribution in
the produced raw SWCNT material. Several different methods have been applied to gain this information. The methods
comprise local probes as transmission electron microscopy (TEM) \cite{rinzler}, scanning tunneling microscopy
combined with tunneling spectroscopy \cite{stm-sts}, and  bulk sensitive probes, such as Raman
scattering\cite{hans,kuzmany98epl}, optical absorption spectroscopy \cite{jost,hk}, electron diffraction
\cite{TEM_diff,TEM_diff1}, X-ray diffraction (XRD)\cite{rinzler}, and neutron scattering \cite{rols,nd}. In the
following, we briefly compare these methods.

\par
On a local scale the distribution of SWCNT diameters has been analyzed using TEM \cite{rinzler,thess}. From these
studies, for material synthesized using laser ablation (in general) a Gaussian diameter distribution was observed,
whereas from other processes such as CVD and HPCO, the SWCNT diameters are found to be spread over a wider range
without a simple distribution function \cite{co}. The big disadvantage of TEM in this context is that it is
nanoscale and thus that one can never be sure that one has obtained a result truly representative of the bulk
SWCNT diameter distribution. Consequently, a number of bulk sensitive methods which provide information regarding
the diameter distribution have been applied, such as electron diffraction and XRD, also neutron diffraction. Since
the SWCNT produced by laser ablation, carbon arc and HPCO are predominantly organized in bundles, X-ray
\cite{rinzler,hans} and electron diffraction are suited to characterize the SWCNT mean diameter and diameter
distribution from the diffraction pattern of the bundle lattice. Such diffraction-based methods possess the
disadvantage that they are insensitive to any individual (and hence non-bundled) SWCNT present in the sample. This
is a grave set-back in the analysis of material from the CVD route, as this material generally contains a
significant proportion of individual nanotubes.

\par

The second, bulk sensitive characterization method utilizes the optical properties of the SWCNT. From
consideration of the folding of the Brillouin zone of a graphene sheet resulting from the new boundary conditions
generated by the conceptual 'rolling up' to form a nanotube, tight binding (TB) calculations have led to a simple
relationship between the SWCNT diameter and the energy of the optical interband transitions of the SWCNT
\cite{book}. From the same type of calculations one third of the possible SWCNTs - those with wrapping vectors
$(n,m)$ where $n-m = 3l$ $(l = 0,1,2,...)$ - are predicted to be metallic. All other tubes are semiconductors. The
unit cell of the SWCNTs strongly depends on the choice of $n$ and $m$. The smallest unit cell is found if $n = m$
or either $n$ or $m$ are zero, and in this case is as large as that of the graphene sheet. Such special nanotubes
are called {\it armchair} (chiral angle 30$^\circ$) and {\it zigzag} (chiral angle 0$^\circ$), respectively. All
other tubes are called {\it chiral}. The diameter $d$ of the tubes is related to the components of the Hamada
vector by $d= a_0 \sqrt{m^2+n^2+mn}/\pi$ where $a_0 = 2.46\, $\AA$\;$ is the lattice constant of the graphene
plane. Due to the one-dimensional nature of the SWCNTs, their electronic structure exhibits clear van Hove
singularities. The energetic separation of the pairs of van Hove singularities is inversely proportional to the
tube diameter. The optical response of the SWCNT is dominated by transitions between peaks in the density of
states (DOS) of the valence and conduction bands, with momentum conservation only allowing transitions pairs of
singularities which are symmetrically placed with respect to the Fermi level. Thus, following the van Hove
singularities, the optical transitions in SWCNT are also inversely proportional to the nanotube diameter. For the
first two allowed optical transitions in semiconducting SWCNT, it follows that E$^S_{11}= 2a_0\gamma_0/d$ and
E$^S_{22}= 4a_0\gamma_0/d$, where $\gamma_0$ is  the tight binding nearest neighbor overlap integral. For the
metallic SWCNT, at a first glance the energies of the optical transitions would appear to be proportional to
E$^M_{11}=6a_0\gamma_0/d$. However, more recently, it has been pointed out that the density of states of the
metallic SWCNT is chirality dependent due to the trigonal warp effect, i.e., the energy contours near the Fermi
surface  deviate from a circle\cite{sp1,sp2,sp3,sp4}. This leads to a splitting of the singularities in metallic
tubes, which is maximal for the zigzag variety.

\par
Experimentally, it has been demonstrated that electron energy-loss spectroscopy (EELS) in transmission  and
optical absorption spectroscopy are powerful tools in the study of the mean electronic properties of bulk samples
of SWCNT \cite{jost,hk,topprl}. From both the EELS analysis and the optical absorption data, several distinct
spectral features are observed for energies below 3 eV. These features have been related to the above mentioned
interband transitions between the van Hove singularities in the electronic DOS of the semiconducting and metallic
SWCNT. The inverse proportionality on the nanotube diameter allows a first determination of the mean diameter and
diameter distribution under the assumption of a scaling factor, which is the tight-binding overlap integral
$\gamma_0$ \cite{hans,jost}. In addition, the higher energy resolution available in optical absorption allows the
identification of fine structure within the individual absorption features \cite{jost}. Since this fine structure
is related to individual SWCNT, or groups of SWCNT with similar diameter, the analysis of such data would appear
to offer information as to whether the formation process of SWCNT leads to the existence of preferred wrapping
angles in the nanotube vector map \cite{hamada,jost}.

\par
Complementary to optical spectroscopy, the same inverse proportionality between energy and nanotube diameter is
found in Raman spectroscopy for the so- called radial breathing mode (RBM) of nanotubes which is observed as an
intense features at around 200 cm$^{-1}$. The energy of the RBM scales as $C/d$ where $d$ is the diameter of the
tubes and $C=  234$ cm$^{-1}$nm is a constant determined recently from an {\it ab initio} calculation
\cite{kuerti98prb}. The Raman response for this mode is subject to a strong dependence upon the energy of the
exciting laser used in the Raman experiment. Whereby photo-selective resonance scattering
\cite{kuzmany98epl,hk,rao,bandow,jorio} has been demonstrated to be responsible for both the fine structure in the
RBM lineshape and an oscillatory behavior of the spectral moments \cite{hans}. Consequently, analysis of Raman
data recorded using different laser lines has also been frequently applied as a tool to analyze the SWCNT mean
diameter and diameter distribution in both bulk and nanoscopic samples \cite{bandow,jorio}. However, one has to be
aware that a simple line shape analysis of the Raman response is misleading and the resonance Raman scattering and
the oscillations of the spectral moments have to be included in the detailed analysis \cite{hans}. One remaining
uncertainty in the Raman analysis of SWCNT bundles regards the size of the intertube interaction within a bundle.
The exact strength of this interaction is still unknown, although it is known to lead to a stiffening of the RBM.
From a model calculation using a non-orthogonal tight-binding approximation, a intertube interaction induced
upshift of the RBM of about 8-12 \% was calculated\cite{henrard}. This upshift of the RBM has also used to
estimate the size of the nanotube bundles \cite{kuzmany98epl}.

\par
As the physical properties of SWCNT depend so crucially upon their diameter, it
is an important challenge to arrive at a sound understanding of how their
diameter can be measured in bulk samples. This serves not only our fundamental
understanding of SWCNT as a materials class, but also is a valuable component
in our thinking about SWCNT as a realistic technological material. In this
paper, we present a detailed analysis of the optical properties of laser
ablation-produced SWCNT, with mean nanotube diameters ranging from 0.9 nm to
1.5 nm. We use data from high resolution EELS in transmission together with
optical absorption spectroscopy and electron diffraction, to examine the
relationship between the observed optical transitions and the mean nanotube
diameter. As a first step, the mean diameter of the SWCNT was obtained from
electron diffraction data taken in the EELS spectrometer from large SWCNT
bundles assuming a hexagonal SWCNT lattice. In the next step, a detailed
analysis of the the optical absorption spectra of SWCNT with different mean
diameters is performed within the framework of a tight binding model. The
results show that a best agreement between the simulated and measured optical
absorption is reached when the simulation only includes SWCNT chiralities up to
maximally 15$^\circ$ away from the armchair axis. Finally, the resulting mean
diameter and diameter distribution from the detailed optical analysis of the
nanotubes is compared to the results from the other bulk sensitive methods of
nanotube diameter determination, namely, electron diffraction, X-ray
diffraction, and Raman scattering.

\vspace{0.5cm}

\section{Experimental}

SWCNT with different mean diameters and diameter distributions were produced by laser ablation as described
previously \cite{jost,hk}. Thin films of SWCNT with an effective thickness about 1000 \AA~ were prepared by
dropping an acetone suspension of SWCNT onto KBr single crystals. After the KBr was dissolved in distilled water,
the films were transferred to a standard 200 mesh platinum electron microscopy grid and heated for 6 hours in
ultra high vacuum up to 600$^\circ$C, which has been shown to remove the organic contamination in the SWCNT
films\cite{rinzler}. The EELS measurements were carried out using a purpose-built 170 keV spectrometer
\cite{eels}. The energy and momentum resolution were chosen to be 180 meV and 0.03 \AA$^{-1}$ for the low energy
loss function and electron diffraction, respectively. The optical absorption data were measured on the same
samples as used for EELS using a Bruker 88 covering from the near-infrared region to the ultraviolet with a
spectral resolution of 2 cm$^{-1}$ (0.25 meV). All experiments were performed at room temperature and the EELS
measurements are carried out under ultra-high vacuum conditions.

\par

\section {Results and discussion}

\subsection {Mean diameter from electron diffraction}

During the formation process in the laser ablation oven van der Waals forces
lead to the formation of bundles of SWCNT in which the individual nanotubes are
arranged within a hexagonal lattice. These bundles usually consist of SWCNT
with a finite diameter distribution \cite{TEM_diff}, are believed
to be representative of the mean diameter of bulk samples. Consequently, as
mentioned above, the diffraction pattern of the bundle lattice can be used to
obtain a first estimation of the SWCNT mean diameter, as the intertube distance
is mainly dependent on the nanotube diameter.
\par
We are able to carry out electron diffraction in the EELS spectrometer by setting the energy-loss to zero. Figure
1 shows the raw electron diffraction data from SWCNT samples with six different nanotube mean diameters. We label
these samples A to F, and they contain SWCNT with diameters covering a total range of 0.2 \AA$^{- 1}$ to 6.0
\AA$^{-1}$. The strong increase in intensity in the raw diffraction data at low $q$ seen in Fig. 1 is small angle
scattering originating from large objects in the sample, e.g.,  catalyst particles, amorphous carbon, etc.
Generally, the peaks in a SWCNT diffraction profile can be divided into two parts. The low-{\it q } part of the
diffraction pattern (below 2 \AA$^{-1}$) is only sensitive to the crystalline order in the bundle, thus
corresponding to SWCNT bundle diffraction. The high-{\it q } range (above 2 \AA$^{- 1}$) is sensitive to the
internal structure of individual tubes, whereby the broad peaks near 2.9 \AA$^{-1}$ and 5 \AA$^{-1}$ originate
from the (1 0 0) and (1 1 0) graphite in-plane reflections, respectively \cite{nd-nt}. The interplane reflection
(0 0 2), i.e., the peak near 2 \AA$^{-1}$ is very weak. The absence of a peak corresponding to the (0 0 4)
reflection proves the absence of multiwall carbon nanotubes in the samples \cite{rols}. In the context of these
experiments, all peaks coming from the bundle diffraction can be used to estimate the mean diameter of the
nanotubes. The first-order diffraction peak (1 0) near 0.4-0.5 \AA$^{-1}$ from the hexagonal bundle lattice has
the highest intensity. As can be seen from the dashed arrow in the inset to Fig. 1, there is a strong upshift in
the (1 0) feature on going from sample A to F, indicating a decrease of the mean nanotube diameter.

\par

With electron diffraction, it is possible to determine the average lattice parameter of the two-dimensional
triangular packing in the bundle ropes by measuring the position of the (1 0) Bragg reflection \cite{rols,lambin}.
Assuming a perfect hexagonal lattice, the intertube distance is equal to the tube diameter plus the two times the
van der Waals radius (ca. 0.335 nm). The resultant values for the mean nanotube diameter for the six samples A, B,
C, D, E, and F are 1.46 nm, 1.37 nm, 1.34 nm, 1.30 nm, 1.08 nm, and 0.91 nm, respectively. We note, following from
theoretical consideration \cite{rols}, that this simple analysis is only correct for bundle sizes larger than 20
nm.

\par
Furthermore, once the mean diameter has been arrived at, the expected positions of the high-order diffraction
peaks can also be calculated using the hexagonal lattice model and compared to the experimental data. If the first
bundle peak is weak, its position can also be extrapolated from the positions of the higher-order peaks. In Fig. 2
the position of the first three diffraction peaks -- i.e., (1 0), (1 1), and (2 1) -- are plotted versus the mean
diameter (derived from the position of the (1 0) peak) for the six different nanotube samples. In each case the
solid lines depict the calculated peak position for the ideal hexagonal structure. As the first peak was used as
the calibrated standard, it naturally lies on the theoretical line. For the high-order peaks, there are only small
deviations from the predicted behaviour, showing that the hexagonal lattice is a good description of the nanotubes
within the crystalline bundles.

\subsection {Low energy interband transitions}

As mentioned above, the energetic position of the interband transitions between the DOS singularities are
inversely dependent on the diameter of SWCNT \cite{book}. EELS in transmission measured using low momentum
transfers probes the optical limit, thus the low-energy peaks in the loss function are due to collective
excitations caused by these optically allowed transitions \cite{topprl}. Later, analogous results for the low
energy interband transitions were obtained from optical absorption spectroscopy \cite{jost,hk}. It is interesting
to compare the results from these the two different experiments for the same samples. Typical EELS and optical
results for SWCNT with 1.3 nm mean diameter (sample D) are depicted in Fig. 3. The left panel shows the loss
function at a momentum transfer $q=0.1$ \AA $^{-1}$ between 0 and 9 eV which covers the excitations related solely
to the SWCNT $\pi$ electron system. The strong peak at ca. 6 eV is the so-called $\pi$ plasmon, which is the
collective excitation of the SWCNT $\pi$ electrons. The first three loss peaks below 3.0 eV ( i.e., L$^S_{11}$,
L$^S_{22}$, and L$^M_{11}$) are ascribed to interband transition from EELS measurement.
 In the right panel a typical optical absorption spectrum of the same nanotubes is depicted.
The inset shows the absorbance after subtracting the contributions from the
high energy interband transitions. From Fig. 3 it is obvious that the
contributions of the low energy interband transitions are very similar in the
two experiments and can be easily compared. The peaks related to transitions
between the first and second pairs of DOS singularities in semiconducting
nanotubes (designated E$^S_{11}$ and E$^S_{22}$) are observed at about 0.9 and
1.5 eV, whereas the position of the feature due to the transitions between the
first pair of DOS singularities in metallic nanotubes (designated E$^M_{11}$)
is about 2 eV. Here it should be mentioned that since in EELS  we are probing
collective excitations ( proportional to  Im$(-1/\epsilon)$), the peak
positions are slightly upshifted as compared to optical absorption  which is proportional
to the imaginary part of $\epsilon$, i.e., the peak position of L$^S_{11}$ is
always higher than that of E$^S_{11}$. In principle, one could use a
Kramers-Kronig-analysis to derive the absorption data from the loss
function data (see Ref.\protect\onlinecite{topprl}).

\par
In Fig. 4 we show the experimental results for the interband transitions E$^S_{11}$, E$^S_{22}$, and E$^M_{11}$
from the SWCNT samples with six different mean diameters (samples A to F from Fig. 1). The left panel shows the
EELS data and the right panel optical absorption results (from which the high energy background has been
subtracted). The dashed lines in Fig. 4 indicate the mean positions (center of gravity) of the SWCNT interband
transitions for the peaks from the semiconducting SWCNT as well as the metallic SWCNT. It is obvious that the
energy of these transitions depends strongly on the SWCNT diameter. With increasing SWCNT diameter, all the
interband transitions peaks shift to lower energy. Whereas in optical absorption a distinct fine structure is
observed, especially for the very thin SWCNT, in the EELS measurements no fine structure could be observed, simply
due to the lower energy resolution in EELS. Due to the above mentioned slight upshift of the peak positions in
EELS as compared to in optical absorption, the peak positions in the following analysis are always  referred to
those in the optical data unless otherwise stated. For the samples E and F (which have the smallest SWCNT mean
diameters), the fine structure in optical absorption is strongly pronounced for the E$^S_{22}$ and E$^M_{11}$
peaks. This is a natural consequence of the greater energetic separation of the DOS singularities in these
nanotubes, meaning that each sub-spectrum from a particular SWCNT diameter is more easily distinguished from the
signal from neighboring diameters. This is also reproduced by the fit analysis of the peak shapes below. As a
consequence of the pronounced fine structure in these peaks in the optical absorption spectra, the mean energetic
positions for the E$^S_{22}$, and E$^M_{11}$ features can be only extracted with some uncertainty.

\par

It is now interesting to compare the center of gravities of the three low
energy absorption peaks E$^S_{11}$, E$^S_{22}$ and E$^M_{11}$ with the above
mentioned predictions within the tight binding, or TB-model. The results are
shown in Fig. 5 for the six samples as a function of the inverse SWCNT
diameter, which was estimated from the electron diffraction data discussed
earlier. The solid lines are the predictions from the TB-model using an overlap
integral $\gamma_0=3.0$ eV which is well established value \cite{hans,book}. It
is obvious that the E$^S_{22}$ and E$^M_{11}$ peaks show in general a good
agreement with the model, and only display small deviations at smaller SWCNT
diameters, which can be explained by a small decrease of the overlap integral
$\gamma_0$ \cite{hans}. However, for all SWCNT material studied, the first
optical transition is always upshifted compared to the theoretical prediction.
Recently, the Coulomb interaction has been taken into account in the
calculation of the theoretical spectra of SWCNT using a conventionally screened
Hartree-Fock approach with an effective-mass approximation scheme \cite{ando}
and this approach has been used to explain the experimental data
\cite{exciton}. With the inclusion of the Coulomb interaction between
electrons, the optical transition energies between the valence bands and the
corresponding conduction bands shift to the higher energies. This blue shift is
a net result of two opposing effects. On the one hand electron correlation
enlarges the band gap and on the other hand excitonic effects would be expected
to lead to a red shift of the first absorption feature. The second optical
transition energy, however, scarcely shifts on switching on the correlation as
the two competing effects described above appear to cancel each other almost
exactly.

\par
Therefore, bearing these facts in mind, the energetic position of the second
absorption peak provides a better measure of the SWCNT diameter when analyzing
the data within the framework of the TB-model (in which correlation effects are
not fully accounted for). Interestingly, the impact of the Coulomb interaction
on the E$^S_{11}$ peak is also strongly dependent upon the SWCNT diameter.
Considering the fact that the E$^S_{22}$ peak (which shows little net result of
correlation effects) should occur at an energy twice that of the E$^S_{11}$
peak, the impact of the Coulomb interaction effects can then be easily
visualized by looking at the diameter dependence of the average value of the
energy positions of the centers of gravity of E$^S_{11}$ and E$^S_{22}$.
In Fig. 6, a summary of these effects
for all the measured nanotube samples is plotted as a function of SWCNT mean
diameter. It can be clearly seen that for the fatter SWCNT the effects are
smaller than for the very thin SWCNT. This
diameter dependence cannot solely be explained by a slightly reduced TB overlap
integral for very thin SWCNT ($d<1.1$ nm), and confirms that additional effects
going beyond the one electron TB model have to be taken into account.
Nevertheless, having said that, a detailed analysis of the E$^S_{22}$ and
E$^M_{11}$ peaks still allows a very accurate determination of the SWCNT mean
diameter and diameter distribution, and can even provide indications of whether
their is a chirality dependence in the SWCNT production process.

\subsection {Detailed analysis of the optical absorption}

As mentioned above, the energetic position of the absorption peaks of SWCNT are
proportional to the overlap integral and inversely proportional to the diameter
of the SWCNT \cite{book}. Since the bulk samples consist of a distribution of
SWCNT with mean different diameters and chirality, a pronounced fine structure
corresponding to groups of SWCNT is observed in the optical absorption spectra.
The profile of the E$^S_{22}$ and E$^M_{11}$ features in the SWCNT spectra
provides suitable data from which to determine the mean diameter and the
diameter distribution of the investigated SWCNT from a direct simulation of the
absorption spectra after subtracting the background. Under the assumptions
described below, this approach contains only the SWCNT mean diameter
and diameter distribution as freely adjustable parameters.
The assumptions underlying the analysis routine are:

\par
\vspace*{3mm} a) If present in the sample, all the SWCNT in the vector map give
the same contribution to the overall optical absorption. This is tantamount to
saying that  the transition matrix element is independent of the SWCNT
chirality or diameter.\\
b) The absorption intensity is dominated by transitions between pairs of
corresponding van Hove singularities in the SWCNT DOS (e.g. E$_{11}$ and not
E$_{12}$), and the broadening due, for example, to life-time effects is also
independent of chirality and diameter. \\
c) The SWCNT in the sample have a Gaussian distribution of
diameters.\\
d) $\gamma_0$ is independent of the chirality or diameter. \vspace*{3mm}
\par

Given these points, the corresponding absorption intensity from SWCNTs with
Hamada vector $(n,m)$ and diameter $d(n,m)$ is modulated by a Gaussian
function. The absorption profile of the bulk SWCNT sample can be written as

\begin{equation}
I(E)=f*\sum_{n,m} \exp[\frac{ -(d_{n,m}-d_0)^2}{ 2(\Delta
d)^2}]\frac{ w}{ (E-E_{ii})^2+(w/2)^2}
\end{equation}

 \noindent where $f$ is an overall
scaling factor and $w$ (ca. 40 meV) describes broadening of each single
transition due to the finite lifetime of the band-to band transition and the
finite resolution of the spectrometer. The energy positions $E_{ii}$ ($i$=1, 2,
3 ) are taken from the separation between the maxima of the van Hove
singularities in the SWCNT electronic DOS.
Recently, it has been pointed out that the detailed form of the van
Hove singularities in the one-dimensional electronic density of states of SWCNT
are chirality-dependent. The deviations from a circle in the energy contours
near the Fermi points produce a splitting of the DOS singularities in metallic
nanotubes (the so-called trigonal warp effect)\cite{sp1}. The magnitude of this
effect depends on the chiral angle of the carbon nanotube and is maximal for
metallic zigzag nanotubes and zero for armchair nanotubes
\cite{sp1,sp2,sp3,sp4}.  An approximate analytical expression for the
density-of-states singularities in single-walled carbon nanotubes has been
derived\cite{sp4}, including the energy splitting for an arbitrary chiral angle
in metallic nanotubes. From the work of Ref. \onlinecite{sp4} semiconducting
nanotubes are shown to fall into two classes and transitions between their van
Hove singularities will have a corresponding energy shift. Since in our
analysis we pick up the value from the maxima of van Hove singularities in the
calculated DOS, this effect is implicitly included.  In this way, the experimental
results can be fitted by varying the mean diameter $d_0$ and the
diameter distribution $\Delta d$ in equation [1].

\par

Fig. 7 illustrates the results of such an individual fit to the first three
absorption peaks of SWCNT sample B. In this case, the fitting has been
performed including all the $(n,m)$ pairs in the SWCNT vector map and took a
$\gamma_0$ value of 3.0 eV. The solid line indicates the as-measured data after
background subtraction, and the dotted line is the result of the fit. For the
E$^S_{11}$ peak, the above mentioned blue-shift has been taken into account (in
the form of a somewhat larger $\gamma_0$ value). For this sample we arrive at a
mean diameter $d$ of 1.37 nm and diameter distribution $\Delta d$ of 0.09 nm,
in good agreement with the electron diffraction results.

\par
Although the gross features of the experimental data are well reproduced by the
fit, there are still small deviations regarding the fine structure. One
possibility in this regard would be that not all nanotubes are created with
equal probability in the production process - there could be SWCNT produced
with a preferred chirality. From a simple treatment of previous optical results
from SWCNT produced by laser ablation \cite{jost}, it has indeed been suggested
that nanotubes formed lie closer to the armchair axis than to the zigzag
direction in the SWCNT vector map. Thus it is natural in our fitting of the
optical data to re-investigate this hypothesis by repeating the analysis of
this high-resolution optical data taking as a basis a preferred selection of
nanotube chiralities. To do this, we divided the vector map into slices of
$5^\circ$ chiral angle and repeated the fit for each slice. In general, a good
agreement is observed for chiral angles close to the armchair axis.
\par
As can be seen from Fig. 8, by far the worst agreement with the experimental
data is reached by only considering SWCNT near to the zigzag axis (chiral
angles between 0- 15$^\circ$). The result is much better when all the possible
nanotubes are included in the fit (0-30$^\circ$). Interestingly, a closer
inspection of the results for the peak derived from transitions between the
second pair of van Hove singularities (Fig. 8b), shows that the quality of the
fit is still further improved when only nanotubes with chiral angles between
$15-30^\circ$ are taken in to account. We note here that this trend as regards
the fit results is fully consistent for all of the optical data considered
here, independent of the mean SWCNT diameter. Consequently, we can conclude
that within the framework of the analysis described here, we have gained
additional evidence that SWCNT are preferentially formed closer to the armchair
rather than zigzag axis during the synthetic process.
However, in consideration of the simplifying assumptions made in the
analytical approach taken here we cannot obtain  information about the
exact distribution of nanotubes across the chiralties from bulk measurement.
Further analysis methods focused on  the individual SWCNT such as
STS-STM\cite{dekker1,stm-arm}, resonant Raman\cite{jorio}, small
area TEM diffraction\cite{TEM_diff,TEM_diff1} are required.
In the context of these data and the fit results it is interesting to compare the
apparent preferential formation of SWCNT with chiral angles between
15-30$^\circ$ with conclusions reached from other chirality sensitive
measurements of the individual SWCNT\cite{TEM_diff1,stm-arm} which all have confirmed that chiralities of
SWCNT produced by laser ablation are close to armchair.

\par

Finally, completing this study of the information regarding SWCNT diameter and diameter and chirality distribution
that can be extracted from bulk optical absorption data, we cross-check the mean nanotube diameter obtained from
the simulation of the optical absorbance with the results from the other bulk, diameter-sensitive methods
mentioned ealier. The results of the comparison of x-ray diffraction, resonance Raman scattering, the fitting of
the optical data, and electron diffraction are depicted in Fig. 9. The x-axis gives the mean diameter as given by
the average over all the bulk SWCNT diameter determination methods. Fig. 9 shows a high degree of consistency
between the methods, with the scattering of the mean diameter values from the average being less than 0.05 nm. The
same holds for the diameter distribution, as shown in the inset of Fig. 9.

\section {Conclusions}

A detailed analysis of the optical properties of SWCNT with different mean
diameters as produced by laser ablation was presented. From a combined study of
optical absorption spectroscopy, high resolution electron energy-loss
spectroscopy in transmission and tight binding calculations, we were able to
accurately determine the mean diameter and diameter distribution of the bulk
nanotube samples studied. In general, the absorption response could be
accurately determined by assuming a Gaussian distribution of SWCNT diameters
and applying the inverse proportionality between the SWCNT diameter and the
energy of the absorption features predicted by the tight binding model. Small
deviations from the TB model are observed for the lowest energy main feature -
E$^S_{11}$ peaks - which are attributed to Coulomb interaction effects. A
detailed fit of the optical absorption spectra allows a determination not only
the mean diameter and diameter distribution but also enables additional insight
to be gained into the possible existence of any chirality dependence during the
SWCNT formation process. The best agreement
between the simulated spectra and experiment is observed upon restricting the
chiral angle of the nanotubes to lie within 15$^\circ$ of the armchair axis.
The mean diameter and diameter distribution resulting from the simulation are
in very good agreement with the values derived from other bulk sensitive
methods such as electron diffraction, X-ray diffraction, and Raman scattering.

\vspace*{5mm}

{\bf{Acknowledgments:}}

We thank the DFG (FI 439/8-1) and the EU (IST-NID-Project SATURN) for funding.
One of us (T.P.) acknowledges financial support from the  \"OAW in form of an
APART fellowship and thanks the FWF P14146 for funding. H.K. acknowledges a
Grant-in-Aid for Scientific Research (A), 13304026 from the Ministry of
Education, Science, Sports and Culture of Japan.

\begin{figure}
\caption{Electron diffraction profiles of SWCNT with different mean diameters from six different samples: A (the
fattest nanotubes) to F (the thinnest nanotubes). The inset shows the spectra enlarged in the region of the bundle
peak.}
\end{figure}

\begin{figure}
\caption{Position of the first three electron diffraction features originating
from the hexagonal SWCNT bundle lattice as a function of SWCNT mean diameter
(full circles). The solid lines are the calculated behaviour for an ideal
hexagonal SWCNT lattice structure as described in the text.}
\end{figure}

\begin{figure}
\caption{(a) Loss function of SWCNT with 1.3 nm mean diameter (sample D)
recorded at q=0.1 \AA $^{-1}$ between 0 and 9 eV.  L$^S_{11}$, L$^S_{22}$, and
L$^M_{11}$ are interband transition in loss function from EELS measurement.
(b) Optical absorption spectra of the same
SWCNT between 0 and 3 eV. The inset shows the absorbance in the range of
E$^S_{11}$, E$^S_{22}$, and E$^M_{11}$ interband transitions after subtraction
of the contributions from the high energy interband transitions.}

\end{figure}

\begin{figure}
\caption{(a) Loss function in the region of the low energy interband
transitions for SWCNT with different mean diameters recorded with q=0.1 \AA
$^{-1}$. (b) Optical absorption spectra (after background subtraction) from
SWCNT with mean diameters as indicated.  }
\end{figure}

\begin{figure}\caption{Observed energy of the three lowest energy interband
transitions estimated from the center of gravity of each peak in the optical
absorption spectrum (full circles) as a function of the inverse mean diameter
(from electron diffraction) of the SWCNT. The solid line shows the results
derived from the commonly-used tight-binding-based formula with $\gamma_0=3.0$
eV and $a_0=0.142$ nm.}
\end{figure}

\begin{figure}
\caption{The difference between the energy of the lowest lying interband transition
(E$^S_{11}$) and half of  that of the second transition (E$^S_{22}$) in SWCNT with
different mean diameters as a function of the SWCNT mean diameter. Deviation
from the value 0.5 is an indication of Coulomb interaction effects, which most
strongly affect the energy position of E$^S_{11}$ (for details see text).}
\end{figure}

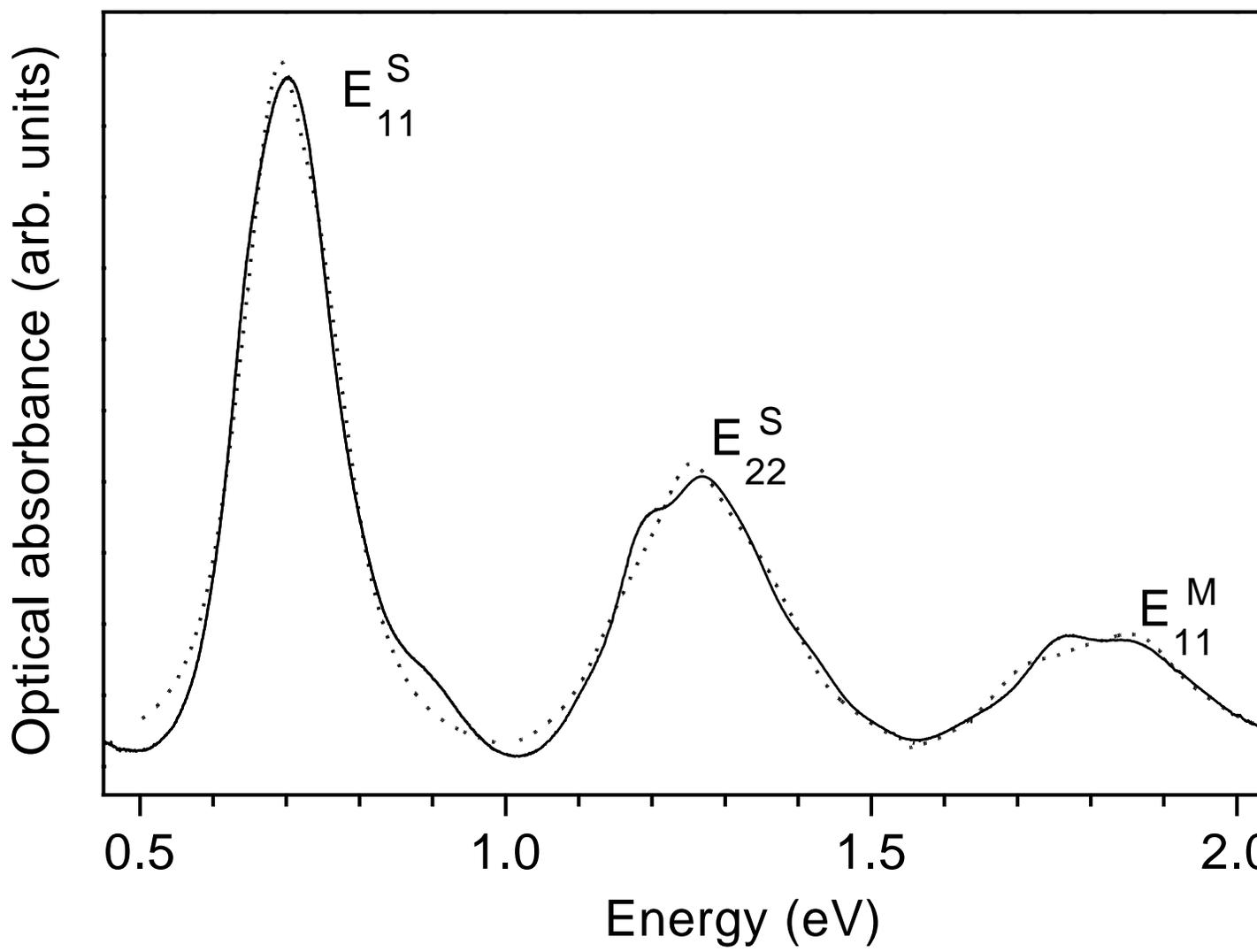
\begin{figure}
\caption{ Simulation of the first three optical absorption peaks of sample B
with d=1.37 nm  upon the basis of tight binding calculations.
Nanotubes of  all chiralities are included. The solid line is the measured
spectrum (with background subtracted); the dashed line represents the results
of the simulation.}
\end{figure}

\begin{figure}
\caption{The tight-binding-based fitting of the (a) E$^S_{22}$ and (b)
E$^M_{11}$ optical absorption features of sample A with d=1.46 nm.
The solid line is the measured data (background subtracted), and the dashed
lines are the simulations using a SWCNT chiral angle distribution as indicated.
30$^\circ$ stands for armchair, 0$^\circ$ for a zigzag nanotube.}
\end{figure}

\begin{figure}
\caption{Correlation plot of the SWCNT mean diameters $d_{method}$ determined by different bulk sensitive methods.
The horizontal axis is the SWCNT $d_{mean}$ averaged over all methods. The Raman and x-ray results are from
Ref.\protect\onlinecite{hans}. The inset shows the plot of the diameter distribution $\Delta$d versus different
mean diameters (full solid), where the open square presents the  data obtained from Raman spectroscopy in
Ref.\protect\onlinecite{hans}.}
\end{figure}

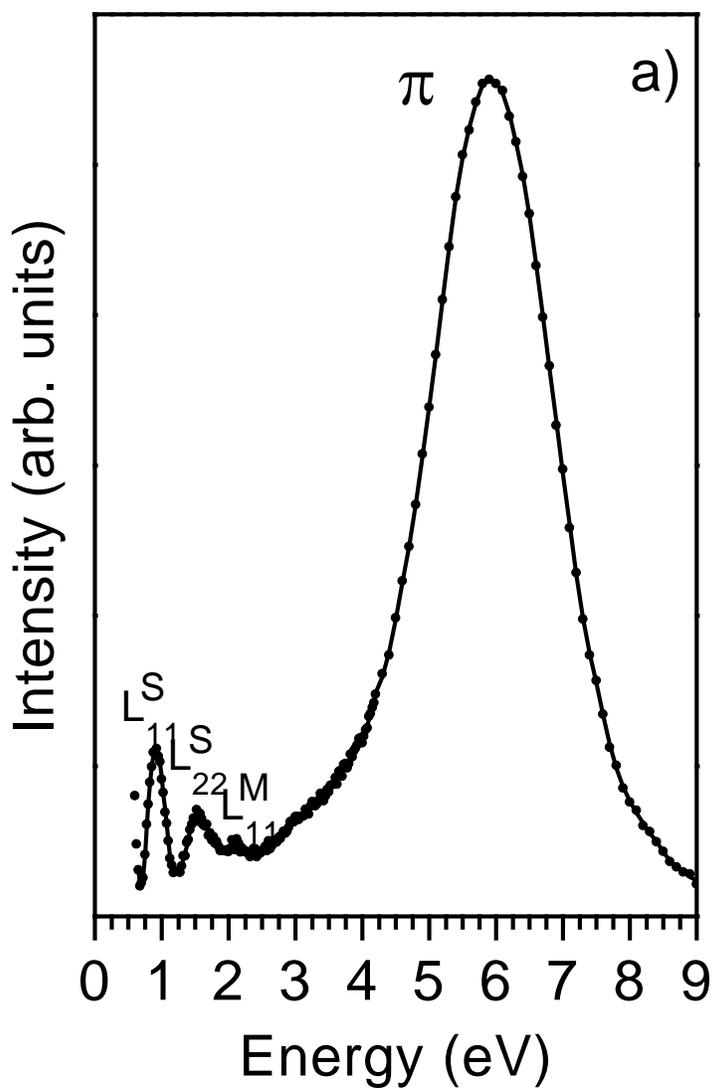
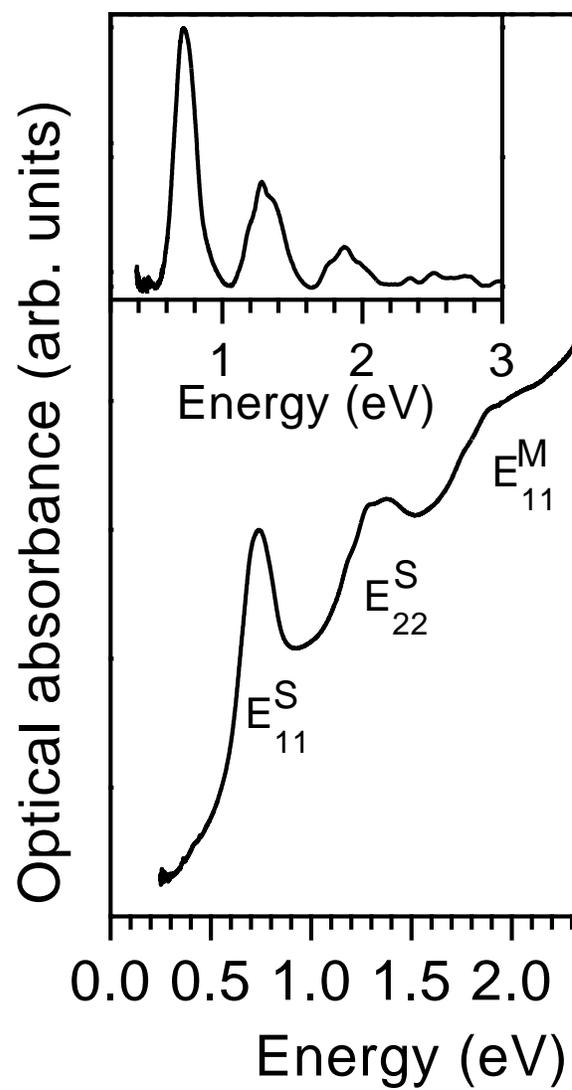
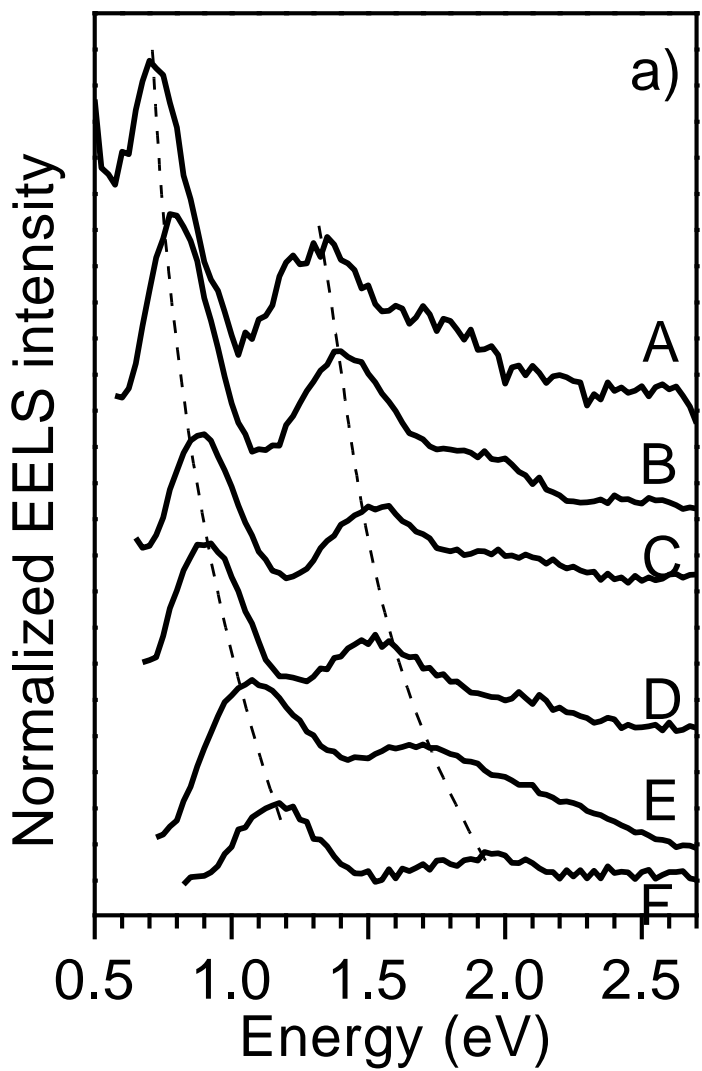
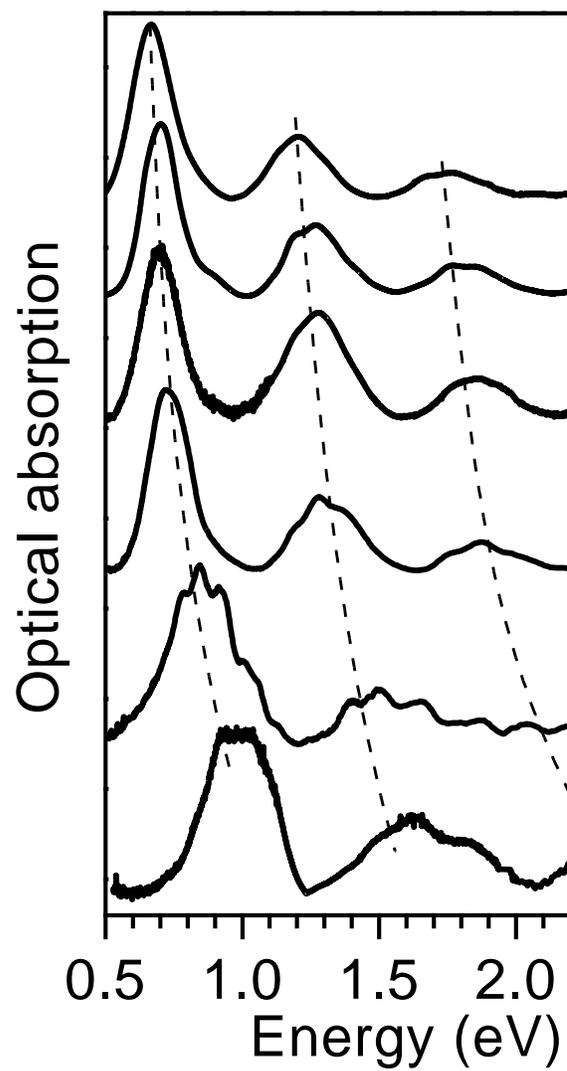
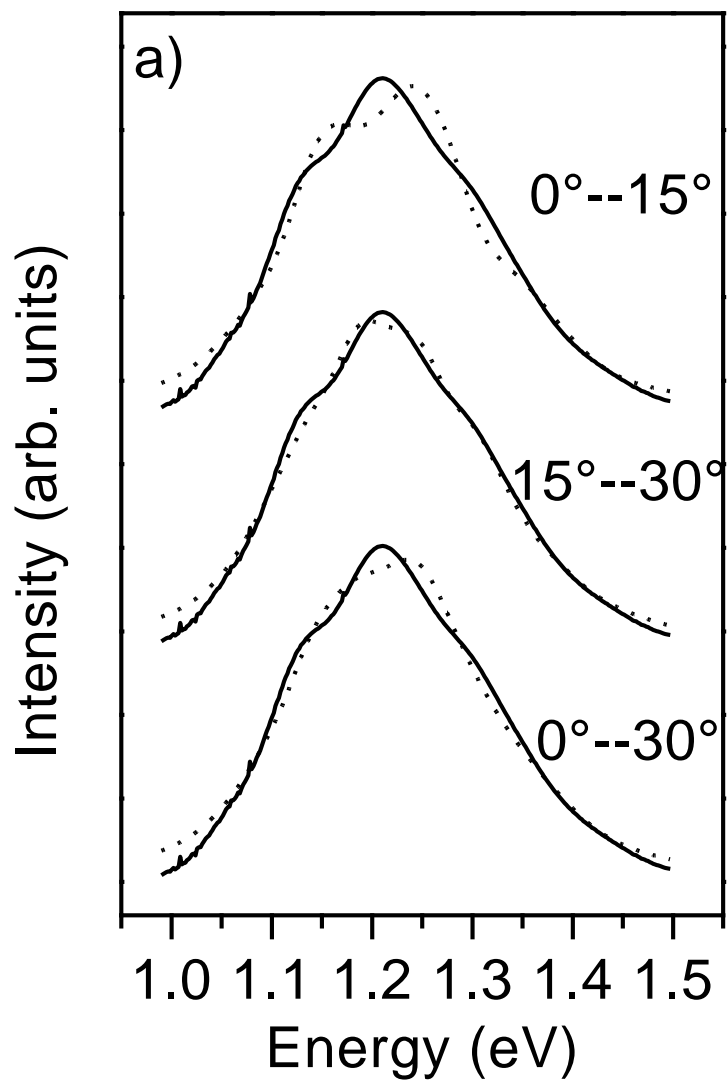
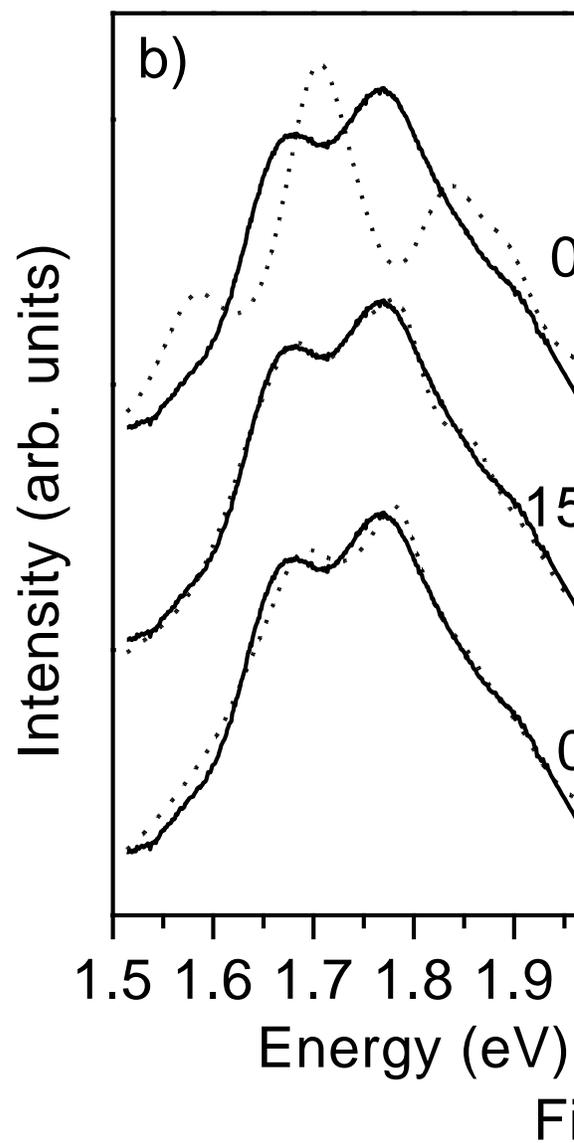
\end{document}